\documentclass[11pt,english]{article}
\usepackage[T1]{fontenc}
\usepackage[latin9]{inputenc}
\usepackage{amsmath}

\makeatletter
\usepackage{amsmath}
\usepackage{amssymb}
\usepackage{latexsym}
\usepackage{epsfig}
\usepackage[numbers,sort&compress]{natbib}
\usepackage{babel}

\newcommand{\be}{\begin{equation}}
\newcommand{\ee}{\end{equation}}

\allowdisplaybreaks

\makeatother

\begin{document}

\vskip 3.0cm

\centerline{\Large \bf Superluminal Neutrinos and  Monopoles}
\vspace*{6.0ex}
\centerline{\large Peng Wang, Houwen Wu and Haitang Yang}
\vspace*{4.0ex}
\centerline{\large \it  School of Physical Electronics}
\centerline{\large \it University of Electronic Science and Technology of China}
\centerline{\large \it Chengdu, 610054, China} \vspace*{1.0ex}
\centerline{pengw@uestc.edu.cn, iverwu@uestc.edu.cn, hyanga@uestc.edu.cn}
\bigskip
\smallskip

\begin{abstract}
In this letter, we show that superluminal neutrinos announced by
OPERA could be explained by the existence of a monopole, which is
left behind after the spontaneous symmetry breaking (SSB) phase
transition of some scalar fields in the universe. We assume the 't
Hooft-Polyakov monopole couples to the neutrinos but not photon
fields. The monopole introduces a different effective metric to
the neutrinos from the one experienced by photons. We find that
the superluminal propagation only exists in a very short distance
from the monopole, about $10^3$ cm in OPERA. No matter how far
they travel, neutrinos always arrive earlier than photons by the
same amount of time, provided a monopole existing on or close to
their trajectories. This conclusion can be tested by future
experiments. The result reconciles the contradiction between OPERA
and supernova neutrinos. We further exclude cosmic strings as a
possible theoretical explanation.
\end{abstract}
\vspace*{10.0ex}

OPERA collaboration \cite{:2011zb} announced their observation
that muon neutrino undergoes superluminal velocity. The flying
time of a beam of $\nu_\mu$, traveling from CERN to the Gran Sasso
laboratory with a baseline distance around $730$km, is $60$ ns
less than that of photons. The result confirms the earlier data
from MINOS \cite{Adamson:2007zzb}. However, the observation
contradicts to the supernova neutrinos from SN 1987A. For its
fundamental impact on the cornerstones of modern physics,
follow-up experiments are certainly demanded. Shortly after the
announcement of OPERA, researchers proposed many theoretical
explanations, respecting or violating Lorentz invariance
\cite{superluminal2011}.
Earlier discussions on superluminal neutrinos can be found in
\cite{earlier superluminal}.


It is well known that after the big bang, spontaneous symmetry
breaking (SSB) phase transition from the false vacua to the true
vacua may not be perfect \cite{Kibble Mechanism}. It is possible
that soliton solutions, also named as topological defects, are
left behind as remnants of the false vacua. The existence of
topological defects may serve as explanations to the neutrino
superluminal propagation. In four dimensional spacetime, there are
three kinds of topological defects: domain walls, cosmic strings
and monopoles, with dimensionality ranging from two to zero
respectively. In a companion paper, we discuss the possible
influences of domain walls on the neutrino superluminal
propagation \cite{Wang:2011sz}. In this letter, we address the
consequences caused by monopoles.

To set up, we consider a Higgs triplet $\varphi^a$, composed of
the adjoint representation of an $SU(2)$ gauge group. In this
letter, the real scalars $\varphi^a$ and gauge fields $B^a$ are
assumed to only couple to neutrinos but not photons. With this
ansatz, the neutrinos see different metrics from a photon does.
This property offers an alternative explanation to the
superluminal behavior of the neutrinos. Since the rest mass of
neutrinos is very small, compared to its energy, it is a good
approximation to assume the neutrinos are massless. The effective
lagrangian is

\begin{eqnarray}
\mathcal{L} & = & -\frac{1}{2}\left(D^{\mu}\varphi\right)^{a}\left(
D_{\mu}\varphi\right)^{a} -\frac{1}{8}\lambda\left(\varphi^{a}\varphi^{a}
-v^{2}\right)^{2}-\frac{1}{4}F^{a\mu\nu}F_{\mu\nu}^{a}\nonumber \\
&  &
+i\overline{\psi}\gamma^{\mu}\partial_{\mu}\psi-\frac{i\,g}{M^{4}}
\overline{\psi}\gamma_{\mu}\partial_{\nu}\psi\left(D^{\mu}\varphi\right)^{a}
\left(D^{\nu}\varphi\right)^{a}\cdots, \label{effective action}
\end{eqnarray}
with
\begin{eqnarray}
\left(D_{\mu}\varphi\right)^{a} & = & \partial_{\mu}\varphi^{a}
+ q\varepsilon^{abc}B_{\mu}^{b}\varphi^{c}\nonumber \\
F_{\mu\nu}^{a} & = & \partial_{\mu}B_{\nu}^{a}-\partial_{\nu}
B_{\mu}^{a}+q\varepsilon^{abc}B_{\mu}^{b}B_{\nu}^{c},
\end{eqnarray}
where the metric on the surface of the earth is approximately set
to the Minkovskian. The first line in eqn. (\ref{effective
action}) is nothing but the Georgi-Glashow model \cite{GSG}. Two
gauge bosons acquire a mass $m_w =q v$ and one remains massless
after the symmetry $SU(2)$ spontaneously broken to a $U(1)$ gauge
group. In the literature, this $U(1)$ field is usually identified
with the electromagnetism $U(1)_{EM}$, since it leads to a charge
double the Dirac charge. However, in our model, we make the
proposal that it is a new unknown field who does not couple to
electromagnetism. Therefore, we call the solitons of eqn.
(\ref{effective action}) as \emph{monopoles} but not magnetic
monopoles. The spinor field $\psi$ stands for neutrinos. The
parameter $q$ is the gauge coupling, different from the electric
charge. The electromagnetism is absent in the Lagrangian since it
does not couple to the other fields. The parameter $M$ denotes the
scale where new physics arises. It is reasonable to believe that
$M\sim m_w$.

With different nonvanishing winding numbers, there are infinitely
many soliton solutions of eqn. (\ref{effective action}). In our
work, we address the simplest case,  the 't Hooft-Polyakov
monopole \cite{Hooft-Polyakov}, with winding number one. The
profiles of the monopole are
\begin{eqnarray}
\varphi^{a}\left(x\right) & = & vf\left(r\right)x^{a}/r\nonumber \\
B_{i}^{a}\left(x\right) & = & a\left(r\right)\varepsilon^{aij}
x_{j}/qr^{2},
\end{eqnarray}
with $\bold x = r(\sin\theta
\cos\phi,\sin\theta\sin\phi,\cos\theta)$. The Latin and Roman
indices denote the three spatial components. The boundary
conditions are $f\left(\infty\right)=a\left(\infty\right)=1$,
$f\left(0\right)=a\left(0\right)=0$, determined by the asymptotic
limits of the fields. From the Lagrangian (\ref{effective
action}), one readily reads the effective kinetic term of
neutrinos

\begin{equation}
i\left[\eta^{\mu\nu}-\frac{g}{M^{4}}\left(D^{\mu}\varphi
\right)^{a}\left(
D^{\nu}\varphi\right)^{a}\right]\overline{\psi}\gamma_{\mu}
\partial_{\nu}\psi.
\end{equation}
With the help of $\partial_{i}\left(x_{a}/r\right)=\left(r^{2}
\delta_{ai}-x_{a}x_{i}\right)/r^{3}$ and $\hat{x}_i=x_i/r$, it is
straightforward to show

\begin{equation}
\left(D_{i}\varphi\right)^{a}\left(D_{j}\varphi\right)^{a}
=\frac{v^{2}}{r^{2}}\left[\left(1-a\right)^2 (\delta_{ij} - \hat
x_i \hat x_j) f^{2} +r^{2}f^{\prime \, 2}\hat x_i \hat x_j\right].
\end{equation}

\noindent Therefore the effective metric neutrinos see is

\begin{equation}
ds^{2} =  -dt^{2}+\left(\delta_{ij} - \frac{g}{M^{4}}\frac{v^{2}}
{r^{2}}\left[\left(1-a\right)^2 (\delta_{ij} -\hat x_i \hat x_j)
f^{2} +r^{2}f^{\prime \, 2}\hat x_i \hat x_j\right]\right)dx^i
dx^j,
\end{equation}

\noindent while the photons still live in Minkovski spacetime. Let
us consider a neutrino travelling along a straight line from $r_i$
to $r_f$, the distances to a monopole. The vertical distance of
the line to the monopole is $R$. The flying time of the neutrino
is
\begin{equation}
t_\nu = \frac{1}{m_w} \Big(\int_{\rho_0}^{\rho_f} \pm
\int_{\rho_0}^{\rho_i} \Big) \frac{\rho\, d\rho} {\sqrt {\rho^2 -
\rho_0^2}}\, \sqrt{1 - \frac{\kappa}{\rho^2} \Big[
(1-a)^2f^2(\rho) \frac{\rho_0^2}{\rho^2} + f'(\rho)^2
(\rho^2-\rho_0^2)\Big]}, \label{neutrionT}
\end{equation}
\noindent where
\begin{equation}
\kappa\equiv \frac{g}{q^2}\left(\frac{m_w} {M}\right)^4,
\hspace{7mm} \rho\equiv m_w r,\hspace{7mm} \rho_0\equiv m_w R.
\end{equation}
Setting  $g\sim q\sim {\cal O}(1)$ is sensible. Therefore, the
dimensionless number $\kappa \sim {\cal O}(1)$. It is hard to
believe that $m_w$, the mass of the vector bosons, is far smaller
than the weak scale $\sim 10^2$ GeV. Then one should note that the
dimensionless parameter $\rho= m_w r$ is a very large number even
for a small distance, for example, 1 cm $\sim 10^{14}/$GeV. Given
the boundary conditions of $f(\rho)$ and $a(\rho)$,
one can readily see that the travelling time difference $\Delta t=
t_c - t_\nu$ between photons and neutrinos is determined by a very
short distance from the monopole, denoted as $\delta$.

In general, there are no closed-form solutions of $f(\rho)$ and
$a(\rho)$. However, in the limit saturating Bogomolny bound, with $\lambda \to 0$,
a BPS soliton solution is available \cite{Srednicki:2007qs}

\begin{eqnarray}
a\left(\rho\right) & = & 1-\frac{\rho}{\sinh\rho}\nonumber \\
f\left(\rho\right) & = & \coth\rho-\frac{1}{\rho}.
\end{eqnarray}
As an illustration, plugging this solution into eqn.
(\ref{neutrionT}), one finds that $\Delta t$ is determined by
$\rho \sim 10$ ($\delta\sim 10^{-15}$cm) distance from the
monopole, up to the precision $10^{-6}$.

Therefore, to account for the superluminal propagation, we assume
there exists one monopole on or very close to the trajectory of
the neutrinos. In this scenario, neutrinos from both OPERA and
supernova SN 1987A arrive earlier than photons by the same amount
of time

\begin{equation}
\Delta t = t_c - t_\nu= \delta - \frac{1}{m_w} \int_0^{m_w \delta}
\sqrt{1- \kappa f^\prime(\rho)^2}\, d\rho \approx 60 {\rm ns},
\label{time diff}
\end{equation}
with a length $\delta \sim 10^3$ cm on the path of neutrinos,
making the real difference on the speed of neutrinos.
If our
conjecture is correct, experiments with different baseline
distances should give the same $60$ ns arrival time difference.

There may exist other possibilities. Specifically, more than one
monopoles are present on the path of the neutrinos. It sounds
reasonable that monopoles are almost evenly distributed in the
earth while few can be met by the supernova neutrinos. If this is
the case, experiments performed in the outer space won't produce obvious
superluminal results and neutrinos on planets always travel faster
than light.



Parallel analysis applies to the one dimensional topological
defect, cosmic strings with a little work. However, the deficit
angle leads to a disaster to nearby objects. Moreover, the lensing
caused by cosmic strings has not been detected yet.

In summary,  we discussed that  a monopole could serve as an
explanation to the recent measurement of superluminal neutrinos in
OPERA. We found that once a 't Hooft-Polyakov monopole is present
on or close to the pathes of the neutrinos, superluminal
propagation arises. Moreover, we showed that the travelling time
difference between neutrinos and photons, caused by a monopole, is
almost fixed no matter how far they travel. From the results of
OPERA, the distance really matters is about $10^3$ cm. This
conclusion explains the contradiction between OPERA and supernova
neutrinos. We also excluded cosmic strings as a possible
explanation due to the deficit angle catastrophe.

\bigskip
\noindent {\bf Acknowledgements} We are grateful to X. Liu for
instructive discussions and inspirations. This work is supported
in part by NSFC (Grant No. 11175039 and 11005016) and Fundamental
Research Funds for the Central Universities (Grant No.
ZYGX2009J044 and ZYGX2009X008).

\end{document}